\documentclass[
hf,
]{ceurart}

\usepackage{listings}
\usepackage{cleveref}
\usepackage{tabularx}
\usepackage[font=normalsize,justification=centering]{caption}
\usepackage[normalem]{ulem}
\begin{document}

%%
%% Rights management information.
%% CC-BY is default license.
\copyrightyear{2023}
\copyrightclause{Copyright for this paper by its authors.
  Use permitted under Creative Commons License Attribution 4.0
  International (CC BY 4.0).}

%%
%% This command is for the conference information
\conference{OM-2023: The 18th International Workshop on Ontology Matching,
  November 7th, 2023, Athens, Greece}

%%
%% The "title" command
\title{Matching Table Metadata with Business Glossaries Using Large Language Models}

% \tnotemark[1]
% \tnotetext[1]{You can use this document as the template for preparing your
  % publication. We recommend using the latest version of the ceurart style.}

%%
%% The "author" command and its associated commands are used to define
%% the authors and their affiliations.
% \author[1,2]{Dmitry S. Kulyabov}[%
% orcid=0000-0002-0877-7063,
% email=kulyabov-ds@rudn.ru,
% url=https://yamadharma.github.io/,
% ]
% \cormark[1]
% \fnmark[1]
% \address[1]{Peoples' Friendship University of Russia (RUDN University),
%   6 Miklukho-Maklaya St, Moscow, 117198, Russian Federation}
% \address[2]{Joint Institute for Nuclear Research,
%   6 Joliot-Curie, Dubna, Moscow region, 141980, Russian Federation}

% \author[3]{Ilaria Tiddi}[%
% orcid=0000-0001-7116-9338,
% email=i.tiddi@vu.nl,
% url=https://kmitd.github.io/ilaria/,
% ]
% \fnmark[1]
% \address[3]{Vrije Universiteit Amsterdam, De Boelelaan 1105, 1081 HV Amsterdam, The Netherlands}

% \author[4]{Manfred Jeusfeld}[%
% orcid=0000-0002-9421-8566,
% email=Manfred.Jeusfeld@acm.org,
% url=http://conceptbase.sourceforge.net/mjf/,
% ]
% \fnmark[1]
% \address[4]{University of Skövde, Högskolevägen 1, 541 28 Skövde, Sweden}

%% Footnotes
% \cortext[1]{Corresponding author.}
% \fntext[1]{These authors contributed equally.}

\author[2]{Elita Lobo}[%
%orcid=0000-0001-6913-3598,
email=loboelita@gmail.com,
%url=https://elitalobo.github.io/,
]
\author[1]{Oktie Hassanzadeh}[%
orcid=0000-0001-5307-9857,
email=hassanzadeh@us.ibm.com,
%url=https://www.oktie.com/,
]

\author[1]{Nhan Pham}[%
%orcid=0000-0001-5307-9857,
email=nhp@ibm.com,
]
% \fnmark[1]
\author[1]{Nandana Mihindukulasooriya}[%
orcid=0000-0003-1707-4842,
email=nandana@ibm.com,
%url=https://www.oktie.com/,
]

\author[1]{Dharmashankar Subramanian}[%
orcid=0000-0002-1990-7740,
email=dharmash@us.ibm.com,
%url=https://www.oktie.com/,
]

\author[1]{Horst Samulowitz}[%
orcid=0000-0002-6780-3217,
email=samulowitz@us.ibm.com,
%url=https://www.oktie.com/,
]
%url=https://www.oktie.com/,

\address[1]{IBM Research, Yorktown Heights, NY, United States}
\address[2]{University of Massachusetts Amherst, MA, United States}

%% Footnotes
\fntext[1]{Work done while at IBM Research.}

%%
%% The abstract is a short summary of the work to be presented in the
%% article.
\begin{abstract}
Enterprises often own large collections of structured data in the form of large databases or an enterprise data lake. Such data collections come with limited metadata and strict access policies that could limit access to the data contents and, therefore, limit the application of classic retrieval and analysis solutions. As a result, there is a need for solutions that can effectively utilize the available metadata. In this paper, we study the problem of matching table metadata to a business glossary containing data labels and descriptions. The resulting matching enables the use of an available or curated business glossary for retrieval and analysis without or before requesting access to the data contents. One solution to this problem is to use manually-defined rules or similarity measures on column names and glossary descriptions (or their vector embeddings) to find the closest match. However, such approaches need to be tuned through manual labeling and cannot handle many business glossaries that contain a combination of simple as well as complex and long descriptions. In this work, we leverage the power of large language models (LLMs) to design generic matching methods that do not require manual tuning and can identify complex relations between column names and glossaries. We propose methods that utilize LLMs in two ways: a) by generating additional context for column names that can aid with matching b) by using LLMs to directly infer if there is a relation between column names and glossary descriptions. Our preliminary experimental results show the effectiveness of our proposed methods.

%Column to Business Glossary Matching entails finding the closest glossary item for a given database column name from a business glossary provided at test time. This problem finds its application in data management tasks that require efficient retrieval of relevant tables and columns from large databases. Several works either use stringent n-grams-based pattern-matching formulas or cosine similarity metrics on embeddings of column names and glossary descriptions to obtain the closest match. However, often the business glossaries contain complex and long descriptions and, therefore, need more involved pattern-matching mechanisms to be matched to the right column name. 
%To address this problem, we leverage the power of large language models to obtain more accurate matching. LLMs are trained on large corpora of data and are, thus, capable of identifying more complex relations between column names and glossaries. In this paper, we propose methods that utilize LLMs in two ways a) by generating additional context for column names that can aid with matching b) by using LLMs to directly infer if there is a relation between column names and glossary descriptions. We also design special LLM prompts to improve the efficacy of these methods. 
%Our preliminary experimental results show the potential of our proposed methods.

\end{abstract}

%%
%% Keywords. The author(s) should pick words that accurately describe
%% the work being presented. Separate the keywords with commas.
% \begin{keywords}
%   Business Glossary Matching \sep
%   Tabular Data Analysis \sep
%   Data Lakes
% \end{keywords}

%%
%% This command processes the author and affiliation, and title
%% information and builds the first part of the formatted document.
\maketitle

\section{Introduction}\label{sec:intro}

Large collections of structured tabular data that businesses possess can be invaluable resources for various analytic tasks. Traditionally, such data collections are gathered in large databases or data warehouses, along with mechanisms of collecting and maintaining metadata with well-curated schemas, data catalogs, or master data as a part of a master data management solution. In practice, the overhead of maintaining accurate metadata may be prohibitively difficult and expensive. More recently, enterprises are moving toward collecting all their data in data lakes without any requirements or strict enforcement of metadata availability or quality. As a result, there is a need for solutions that can effectively use limited metadata, such as column headers, and automatically generate useful metadata. Most organizations maintain some business glossary~\citep{business_glossary_book_chapter} with a set of concepts that are relevant to the business processes. If table columns can be annotated with business glossary terms, it helps downstream tasks such as data discovery, data integration, or performing advanced analytics. 

\begin{figure}
%\captionsetup{justification=centering}
\centering
\includegraphics[width=\textwidth]{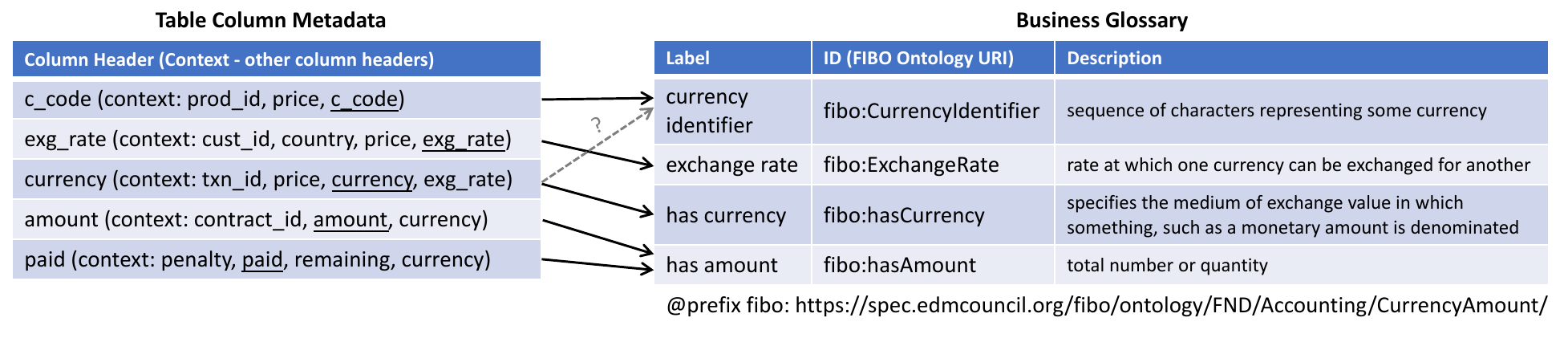}
\vspace{-20pt}
\caption{Example column headers and their associated business glossary terms}
\label{fig:example}
\vspace{-10pt}
\end{figure}

The task of mapping table columns to a business glossary is similar to the task of annotating a table column with an ontology or knowledge graph concept, which is referred to as the Column Type Annotation (CTA) task~\citep{jimenez2020semtab}. However, to our knowledge, prior work has not considered further restricting the task to using only the table metadata (table name and column headers) and business glossary containing labels and descriptions only. The problem we study in this paper is inspired by our ongoing work 
%\np{is this double-blind submission? if yes I wonder this sentence violates the double-blind rule} 
on implementing an automated semantic layer for enterprise data lakes~\citep{Weidele23}, and has the following characteristics: 1) we do not have access to data contents due to access restrictions common in enterprise data lakes; 2) we have tabular data with no metadata other than column headers, which is a result of large data imports from highly heterogeneous sources or automated table extraction pipelines; and 3) there is no or very little training data, as the process of manually labeling table columns with business glossary terms is laborious and requires domain expertise. \Cref{fig:example} shows a few example column headers along with their context (other column headers in the same table) and their associated business glossary terms.

In the absence of rich metadata and an ontology, the matching process can only rely on the header labels and glossary labels and descriptions. Essentially, the problem becomes a string or text similarity matching problem. Prior work has studied various flavors of such matching methods for record matching in databases~\citep{ChandelHKSS07,HassanzadehSM07} as well as ontology alignment~\citep{CheathamH13}. Such methods rely on either {\em syntactic} matching methods, which rely on common tokens and substrings between the terms that should be matched, or {\em semantic} matching methods, which rely on the availability of a dictionary of terms along with lists of related terms such as synonyms, hypernyms, and hyponyms. In our setup, we often need to match terms with very little syntactic similarity, and we do not have access to a dictionary that could enable semantic matching. Column headers in tabular data are often cryptic terms, and business glossaries use terminology very specific to a particular enterprise. More recent work has proposed the use of vector representations of terms in the form of embeddings~\citep{reimers-2019-sentence-bert}; however, such methods require domain-specific training data and tuning.

In this paper, we propose a novel matching solution that relies on the power of Large Language Models (LLMs) to enable the matching of table columns with glossaries when column headers are not very descriptive, and glossary terms do not have a close syntactic similarity to the column headers, and little or no training data is available. In what follows, we first discuss related work. We then present the problem description, and then we present the details of our solution. In \cref{sec:experiments}, we present the results of our experiments using real-world enterprise data and business glossaries. We end the paper by outlining a number of lessons learned and avenues for future work.

\section{Related Work}\label{sec:related_work}
\label{sec:related}
A core problem in semantic table understanding~\citep{pujara2021tables} is column type annotation, i.e., annotating table columns with a type from an ontology, which enables many business intelligence tasks such as semantic retrieval, data exploration, and knowledge discovery. SemTab challenge~\citep{jimenez2020semtab}, which aims at benchmarking systems dealing with the tabular data to KG matching problem, provides several datasets in which the column type annotation task can be evaluated. In the SemTab challenge, this task is formulated as an unsupervised task where participating systems are not given training data. %Each year, the SemTab challenge introduces new datasets.

\paragraph{Column type annotations using table data} 

MTab~\citep{DBLP:conf/semweb/NguyenYKI021a}, JenTab~\citep{abdelmageed2021jentab}, and DAGOBAH~\citep{huynh2021dagobah}, are an example of the systems that participated in the SemTab challenge comprising three KG matching tasks, namely, cell to KG entity (CEA task), column to KG class (CTA task), and column pair to KG property (CPA task). As these systems typically solve the three tasks in a joint manner, they follow a pipeline architecture. The first step links cell mentions to entities within the target ontology. The second step predicts the most likely type for the query column based on the linking results. MTab and DAGOBAH also use additional information from the graph, such as entity relations, to improve cell linking accuracy. It is a requirement for these systems to have cell values that can be mapped to KG entities, which might not be the case in most industry tables. 
% Other recent work~\citep{DBLP:conf/aaai/ChenJHS19,DBLP:conf/ijcai/ChenJHS19,DBLP:journals/pvldb/DengSL0020} has studied this task based on cell values only, i.e., using only the available information in a table for linking columns without doing entity linking first. 
%oktie: removed below to save space.
%There are also a number of datasets used in this task. 
%Deng et al.~\cite{DBLP:journals/pvldb/DengSL0020} introduce a new dataset for the supervised CTA task by annotating WikiTables columns with Freebase types. Viznet~\cite{viznet} is a large-scale corpus of over 31 million datasets. The dataset of particular interest here is the WebTables dataset. These datasets are often used as a source of training data. \textsc{Sato}~\cite{DBLP:journals/pvldb/ZhangSLHDT20} is a supervised model that combines topic modeling and structured learning with single-column type prediction based on Sherlock~\cite{DBLP:conf/kdd/HulsebosHBZSKDH19}. \textsc{Sato} is trained on a subset of the WebTables dataset comprising relational tables with valid headers only.

\paragraph{Column type annotations with only using metadata} The problem setup that is studied in this paper differs from the traditional column type annotation task. In our setup, the system that is performing the matching between the table column and glossary concepts will only have access to the table metadata (i.e., table name and column headers) but not the actual table data (i.e., cell values). This problem setup has similarities with the ontology matching methods that rely purely on string similarity measures.
%TODO: Do we have some related work to cite here?
%paragraph{String similarity-based matching} As mentioned earlier, 
String similarity measures have been studied extensively for various matching tasks, including in ontology alignment~\citep{CheathamH13}. Syntactic measures of similarity measure how close two strings are based on measuring the overlap between tokens or substrings in two strings or measures based on the number of character edit operations (e.g., removal or replacement) that can transform one string to another. Examples of such methods are edit similarity and Jaro-Winkler~\citep{winkler90}. While such approaches have shown very promising performance in various matching tasks, they are inherently not capable of differentiating between strings that are syntactically very similar but semantically dissimilar. Classic semantic measures rely on resources such as WordNet~\citep{Lin2008} containing related terms. The application of those methods is limited to when such resources are available. More recently, methods that rely on vector representations for semantic similarity have shown superior performance in various tasks. Initial approaches relied on word2vec~\citep{Liao2021}, which can handle semantic matching between words and short phrases. More recently, word embeddings~\citep{wang-etal-2018-ontology,Chen2021embeddings,Zhang2014word,kolyvakis-etal-2018-deepalignment,Tounsi2019Embedding} and sentence embeddings~\citep{reimers-2019-sentence-bert} have shown promising performance in semantic textual similarity tasks. As we will show in our experiments, business glossaries often have very similar labels and descriptions that these sentence transformer-based approaches alone cannot effectively differentiate between.

\section{Preliminaries}
\label{sec:preliminaries}
\subsection{Problem Setup}
We assume a setting in which we have only superficial tabular metadata corresponding to any chosen table in the form of a list of $n$ superficial metadata fields, say $M = \{(M_i)_{i=1}^n\}$. In practice, this list could contain a main column name of interest that must be matched with the correct business concept in a glossary, along with the other column names in the table under question. We also assume access to 
%we have a list of $n$ tabular metadata $\mathcal{M}=\{(M_i)_{i=1}^n\}$ where $M_i$ represents the $i^{th}$ table metadata derived from databases and 
a business glossary that consists of $m$ glossary items $\mathcal{G}=\{(l_j,d_j)\}_{j=1}^m$. Here, $l_j$ and $d_j \forall j\in[1,m]$ represent respectively the \emph{label} and \emph{description} of the $j^{th}$ glossary item. For example, the glossary could be a list of tuples containing labels and descriptions of various business concepts. Given such superficial metadata $M$, the task is to find its closest glossary item match.

In this paper, we consider a relaxed version of the glossary matching problem, where the task is to select $k$ glossary items for any given metadata $M$ such that it maximizes the probability of \textbf{Hit@k} for any given $M$. A \textbf{Hit@k} for a given metadata represents a Boolean variable that takes the value \emph{one} if the selected $k$ glossary items contain the closest match of the metadata and \emph{zero} otherwise.
Finally, we also assume that we have a human feedback bank available in the form of $l$ tuples  $\mathcal{H}=\{(M_k, G_k)_{k=1}^l\}$  where $M_k$ represents some metadata and $G_k\in\mathcal{G}$ represents the correct glossary match. We will use the human feedback bank $\mathcal{H}$ to construct task demonstrations for the In-Context Learning approach described in the next section.

% If no such glossary item exists, we expect the algorithm to return "None of the above"

\subsection{Large Language Models}
Recent work \citep{Ouyang2022TrainingLM,Wei2022Finetuned,zhou2023large,jang2022can,Zhao2023ASO} has demonstrated that Large Language Models (LLMs) perform extraordinarily well on instruction-based tasks as long as these tasks can be represented in natural language. 
 LLMs are transformer models with billions of parameters, trained on large data corpora and fine-tuned on instructions-based tasks, including classification tasks, generation tasks, and question-answering tasks. 
An LLM takes as input a prompt containing the description of the task along with additional context represented using natural language and outputs the results of the task in natural language.  
In this work, we leverage LLMs, specifically, Flan-t5-models \citep{longpre2023flan} to obtain more accurate metadata for business glossary matching.
Since LLMs have been trained on large data corpora, they can identify complex relations and patterns between different objects in natural language and, thereby, can be used to obtain more accurate matching.

\subsection{In-Context Learning}
\begin{figure}
\captionsetup{justification=centering}
\centering
\includegraphics[page=3,width=\textwidth]{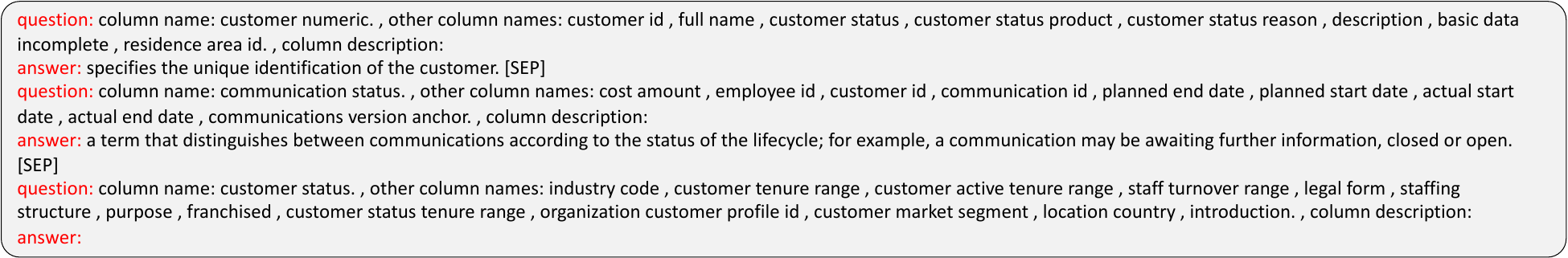}
\vspace{-10pt}
\caption{Example prompt 
for tuned and pre-tuned Flan-t5 models 
used in MDG-MICL (2-shots) method.}
\label{fig:mdg_micl}
\vspace{-10pt}
\end{figure}

Fine-tuning LLMs for new tasks or datasets is often very computationally expensive and requires large amounts of data, which is often not feasible. A common approach to evade this problem is by appending one (one-shot) or multiple (multi-shot) demonstrations of the task in natural language to the prompt. This is commonly known as the One-shot or Multi-shot In-Context Learning (ICL) or In-Context Prompting method~\citep{min2022rethinking}. \Cref{fig:mdg_micl} shows an example of multi-shot in-context learning on a classification task. In-context learning is known to have worked well for several new problems in the prior literature. This work uses the human feedback bank to generate relevant demonstrations for in-context learning. We conjecture that ICL, with good demonstrations, can improve the performance of the glossary matching task without additional fine-tuning.

\section{Methodology}\label{sec:methodology}
\label{sec:methods}
We propose two different classes of methods a) Metadata Description to Glossary Matching (MDGM) b) Direct Metadata to Glossary Matching (DMGM) that retrieve a set of $k$ glossary items for any given metadata such that it contains the glossary match with high probability. In MDGM methods, we use the Large Language Models to obtain a metadata description and use the description to retrieve $k$ glossary items that are most similar to the given description in some latent space. 
On the other hand, in the DMGM methods, we use LLMs to directly match metadata to business glossaries. More specifically, we treat the metadata to glossary matching problem as either a Boolean classification or a Multi-class classification problem. We design special prompts to LLMs to \emph{directly} infer which glossary items are potential descriptions of the metadata and choose the top-$k$ glossary items most likely to be the description of the given metadata.
Although MDGM methods seem like an indirect approach to glossary matching, they can be useful when the glossary constantly changes during test time, and direct inference over large glossaries is expensive.
We show in \cref{sec:results} that MDGM methods tend to outperform DMGM methods.

\subsection{Column Description to Glossary Matching}
We now describe various techniques we propose for generating descriptions of metadata for the metadata to business glossary matching problem.

\subsubsection{Metadata Description Generation via Multi-Shot In-Context Learning (MDG-MICL)}
Since LLMs are trained on large data corpora, they can generate good descriptions of any concept they may have seen during the training period. In this method, we leverage this knowledge of LLM to generate descriptions of the given metadata. We design a special prompt instructing the LLM to generate a metadata description. Further, we use ICL to improve the quality and control the format of the description generated by the LLM. \Cref{fig:mdg_micl} shows an example of the ICL prompt used for this task with tuned Flan-t5-xl and Flan-t5-xxl models. 
%\np{Figure 2 is mentioned here but displayed very far below, maybe bring it closer to here.} 
%oktie: moved the figure
To construct demonstrations for ICL, we proceed as follows. Using the Sentence-BERT (SBERT) sentence embeddings~\citep{reimers-2019-sentence-bert}, we first generate sentence embeddings of the metadata and all the descriptions corresponding to the glossary items in the human-feedback bank, $\mathcal{H}$. We use the cosine similarity metric to find $e$ glossary item descriptions from  $\mathcal{H}$ closest to the metadata in the SBERT sentence embedding space.
We construct demonstrations from these $e$ glossary items in $\mathcal{H}$, and append them to the prompt, in response to which the LLM generates a description. We obtain the final description by appending the table name and metadata to the LLM-generated description. We embed this description using SBERT and obtain the top-$k$ glossary items by computing the cosine similarity metric between this embedding and the sentence embeddings of all the glossary item descriptions. 
The procedure of computing top-$k$ glossary items from the glossary set $\mathcal{G}$ for a given metadata description using the SBERT sentence embeddings and cosine similarity metric is used in several subsequent methods; for the sake of brevity, we will here on refer to this procedure as the SBERT $k-$nearest neighbors method.

\subsubsection{Metadata Description Generation via Classification (MDG-Cl)}
As discussed in \cref{sec:preliminaries}, LLMs can also identify complex relations between various concepts in natural language. Therefore, in this method, we leverage LLMs to directly select the best description from the set of glossary descriptions in $\mathcal{G}$ using a classification-based technique.

Specifically, for a given metadata and glossary set $\mathcal{G}$, we construct a binary classification prompt against each glossary item in the set $\mathcal{G}$, that queries the LLM on whether the given glossary item is a potential description of the metadata. \Cref{fig:mdg_micl} shows an example of the classification prompt used for this task.
Note that when the glossary set $\mathcal{G}$ is too large, it may result in high inference costs.
We can mitigate these high costs by first shortlisting the top $k_1, (k_1 > k)$ glossary item from the glossary set $\mathcal{G}$ using the SBERT $k-$nearest neighbors method before proceeding with the classification prompts. 

\begin{figure}
\captionsetup{justification=centering}
\centering
\includegraphics[page=2,width=\textwidth]{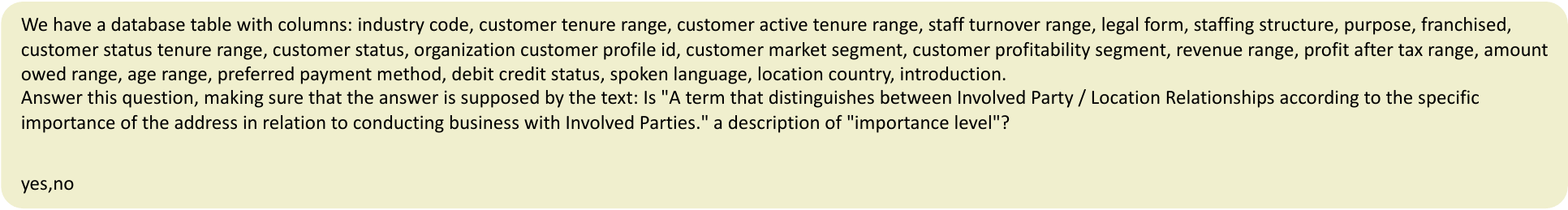}
\caption{Example prompt for tuned Flan-t5-xl and pre-tuned Flan-t5-xxl models used in MDG-Cl method.}
\label{fig:mdg_mcl}
\end{figure}

The final description of the metadata corresponds to the glossary description for which the classification response was positive with the highest log probability score of the classification task, appended to the table name and the metadata itself. If no such glossary item exists, we use the metadata itself as the description. 
Once the metadata description is generated, we select the top-$k$ glossary items from the glossary set $\mathcal{G}$ using SBERT $k-$nearest neighbors described in MDG-MICL method. 
% \np{This sentence is a little disconnected with the previous one} 

\subsubsection{Metadata Description Generation via Multiple Choice Question Answering (MDG-MCQA)}
\begin{figure}
\captionsetup{justification=centering}
\centering
\includegraphics[page=1,width=\textwidth]{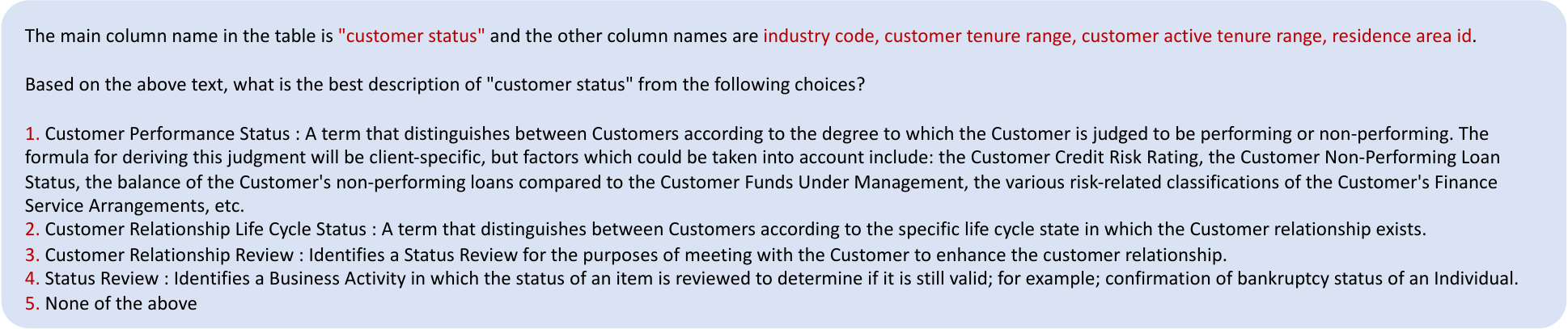}
\caption{Prompt for Flan-t5-xl and pre-tuned Flan-t5-xxl models used in MDG-MCQA method.}
\label{fig:mdg_mcqa}
\end{figure}

An alternative way of generating descriptions using LLMs is by using a Multiple Choice Question Answer (MCQA) prompt, as shown in \Cref{fig:mdg_mcqa}, that instructs the LLM to choose the best description of the metadata amongst the descriptions of selected glossary items. Although this may seem counterintuitive, we observe in our experiments that using the description of the selected glossary item to find the top-$k$ glossary items instead of simply returning the glossary item corresponding to the selected description results in a higher Hit@5 rate.
Similar to previous methods, we shortlist the top $k_1 (k_1>k)$ glossary items that are closest to the metadata in sentence-embedding space and use them as choices in our Multiple Choice Question Answer prompt.
We also add "None of the above" to the list of choices in the MCQA prompt. 
Finally, we append the metadata and the table name to the description of the LLM-selected glossary item and use it to find the top-$k$ glossary items using the SBERT $k-$nearest neighbors method. Note that when the LLM selects the ``None of the above" option, we use the metadata itself as the description. 

\subsection{Direct Metadata to Glossary Matching}

\subsubsection{Direct Inference via Classification (DI-Cl)}

\begin{figure}
\captionsetup{justification=centering}
\centering
\includegraphics[page=5,width=\textwidth]{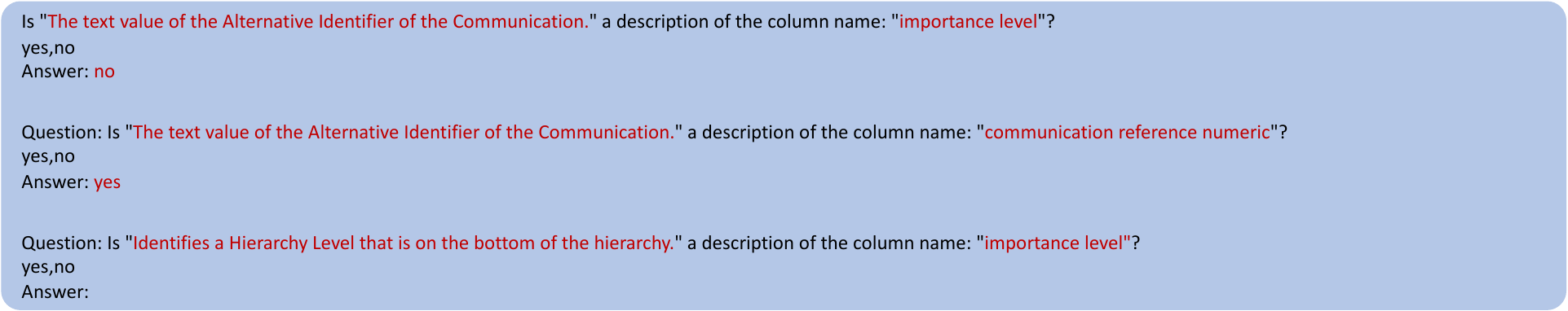}\caption{Example prompt for all pre-tuned and tuned Flan-t5 models used in DI-Cl method.}
\label{fig:di_cl}
\end{figure}

This method is a variant of the MDG-Cl method that uses LLM to directly select the top-$k$ glossary items for any given metadata without needing to generate a description of the metadata. Similar to previous methods, we shortlist the top $k_1 \,(k_1> k)$ glossary items closest to the metadata in the sentence-embedding space. For each of these glossary items, we construct binary classification prompts that query the LLM on whether the description of the glossary item matches the metadata. An example prompt is shown in Figure~\ref{fig:di_cl}. Among all glossary items with positive responses, the glossary items with top-$k$ highest log probability scores are selected. It is important to note that this method may return less than $k$ glossary items. Such missing items are assumed to be incorrect matches while computing the \textbf{Hit@k} metric.

\subsubsection{Direct Inference via Multiple Choice Question Answering (DI-MCQA)}
\begin{figure}
\captionsetup{justification=centering}
\centering
\includegraphics[page=4,width=\textwidth]{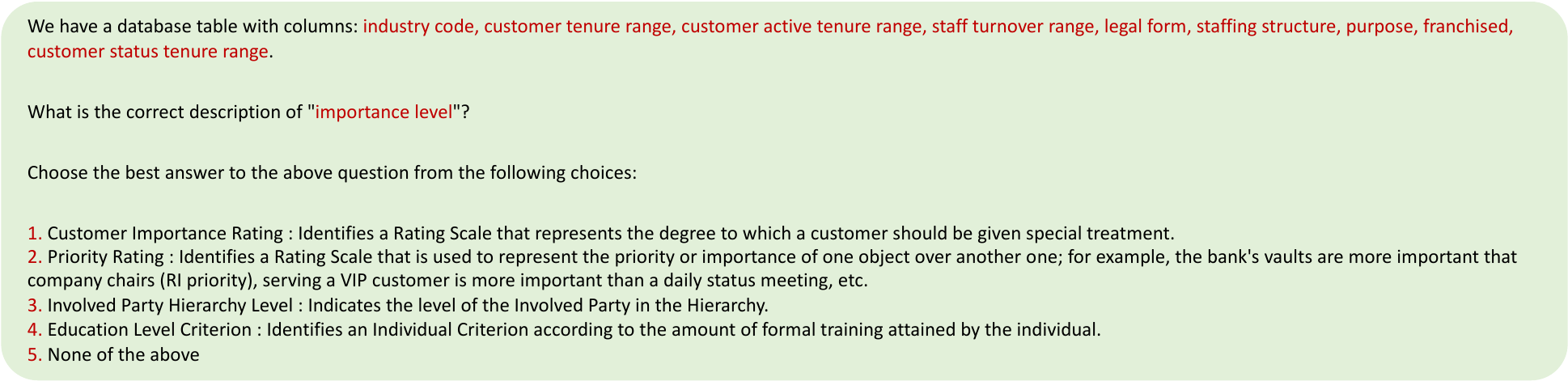}
\caption{Example prompt for Flan-t5-xl and pre-tuned Flan-t5-xxl models used in DI-MCQA method.}
\label{fig:di_mcqa}
\end{figure}

We consider another variation of the MDG-MCQA method that computes a single best match of the given metadata without generating a description of the metadata. In this method, we follow the same procedure as MDG-MCQA to construct Multi-Choice Classification prompts, as shown in Figure~\ref{fig:di_mcqa}, and return the glossary item corresponding to the description selected by the LLM. Since this method always returns a single glossary item, we will assume that the $k-1$ missing glossary items are incorrect matches while computing the \textbf{Hit@k} metric.

\section{Experiments}
\label{sec:experiments}
In this section, we empirically evaluate and compare the performance of all the MDGM and DMGM methods proposed in \cref{sec:methodology}. Specifically, we investigate the following two questions a) Which of the proposed methods: MDG-MICL (0-shot, 1-shot, 2-shots), MDG-Cl, MDG-MCQA, DI-Cl, and DI-MCQA,  best leverage LLM in solving the metadata to glossary matching task, and b) is it possible to obtain more accurate matching, i.e., higher \textbf{Hit@5} and \textbf{Hit@1} using LLMs than using basic similarity-score based matching methods? Our preliminary results show that LLMs are indeed effective in improving glossary matching accuracy. 

\subsection{Experimental setup}
In all our experiments, the metadata consists of the column name of interest and the other column names in the table. The glossary is a list of tuples where each tuple consists of a label and a description of the label. Multiple column names may be matched to the same label. Furthermore, the label of the matched glossary item for each column name may not coincide with the column name. 
We evaluate our methods on the Flan-T5-XL, Flan-T5-XXL \citep{chung2022scaling,longpre2023flan}, and a Flan-T5-XL model that we fine-tune on the training dataset using the supervised fine-tuning method for LLMs \citep{ouyang2022training}.
For each method and LLM model, we experiment with 4-6 different prompt templates. However, due to lack of space, we only provide examples of the prompt templates that achieved the highest hit@5 rate (\ref{fig:mdg_micl},\ref{fig:mdg_mcl},\ref{fig:mdg_mcqa},\ref{fig:di_cl},\ref{fig:di_mcqa}).

We use the \texttt{all-mpnet-base-v2} SBERT model from the \texttt{sentence-transformers} library\citep{reimers-2019-sentence-bert} in all experiments. 
We evaluate each method based on their \textbf{Hit@5} and \textbf{Hit@1} rates on the test dataset, i.e., the empirical mean of \textbf{Hit@5} and \textbf{Hit@1} computed on the test dataset. These measures were chosen to reflect our goal of having the correct glossary item as the top or within the top 5 glossary items returned to the user on a GUI~\citep{Weidele23}.
% In experiments that select glossary items based on some scores, for example, the MDGM methods and DI-Cl method, we find the best threshold 
We compare these scores against ones produced by a baseline, which computes the top-$k$ glossary items based on the cosine similarity of the sentence embedding between the column name and the descriptions in the glossary. 

% \subsection{Datasets}

\paragraph{MDE Dataset}
The MDE Dataset is an IBM-internal benchmark developed by annotating the column names of the ``Customer Insight'' example database of ``IBM InfoSphere Warehouse Pack"~\citep{infoSphere2010} with theevaluateglossary terms from the IBM Knowledge to fine-tune Financial Services (IBM KAFS)~\citep{kafs}
%\footnote{\url{https://dataplatform.cloud.ibm.com/docs/content/kaaas/User_Guide/pro_ove/ika/kafs.html}} 
glossary. The Customer Insight database consists of 26 tables with 688 columns. The column names contain cryptic codes and abbreviations to reflect realistic tables commonly seen in client engagements. The IBM KAFS glossary contains 9,137 business terms with their labels and descriptions. Out of 688 columns, 488 have suitable matching terms in the glossary, and the rest are annotated as null mappings and ignored in the evaluation. We split the mappings into train, test, and demonstration splits with 208, 212, and 68 columns, respectively. These splits contain tuples of the form $(column\_name, other\_column\_names,glossary\_item)$. The training, test, and demonstration splits are used for fine-tuning the LLM model, evaluating the method, and as a proxy for human feedback.

% \subsection{Evaluation Metric:}

\subsection{Experimental Results}\label{sec:results}
\Cref{table:pretuned} shows the \textbf{Hit@5} and \textbf{Hit@1} rates achieved by different methods on the MDE dataset with different LLM models. As expected, the \textbf{Hit@5} rate increases with the number of demonstrations in the MDG-MICL method. This result suggests that it may be beneficial to use in-context learning whenever demonstrations are readily available. However, we do not observe a similar trend for \emph{Hit@1} rate. We believe this is due to the difficulty of matching metadata to a single glossary item when multiple glossary items have similar descriptions. Overall, MDG-MICL achieves the highest Hit@5 and \textbf{Hit@1} scores and significantly outperforms the baseline method when two demonstrations are provided in the prompt.
Meanwhile, we observe that the DI-Cl and DI-MCQA methods achieve the worst \textbf{Hit@5} rates. We conjecture that this may be due to the underlying biases of LLMs towards certain class labels in classification and question-answering tasks as observed in prior works \citep{xu2023knn,zhao2023llm}. These biases can be corrected using various calibration techniques \citep{xu2023knn,zhao2023llm}, which we leave for future work. It is also important to note that the DI-MCQA method selects a single best glossary match and, thus, more likely fails to select the correct item when multiple glossary items have similar descriptions. These preliminary results indicate that LLMs alone may not improve the matching accuracy. 
Finally, although MDG-Cl and MDG-MCQA use the same classification prompt and Multiple Choice Question Answer prompts as DI-Cl and DI-MCQA, they achieve higher \textbf{Hit@5} and \textbf{Hit@1} than the latter methods. We believe that this is because these methods select glossary items whose descriptions are similar to the closest glossary match and, thus, tend to perform better.

\begin{table}
    \centering
  \begin{tabular}{ p{3.8cm}p{1.3cm}p{1.3cm}p{1.3cm}p{1.3cm}p{1.3cm}p{1.3cm}p{1.3cm}}
    \toprule
    \multirow{2}{*}{Methods} &
      \multicolumn{2}{c}{Flan-t5-xl } &
      \multicolumn{2}{c}{Flan-t5-xxl } &
      \multicolumn{2}{c}{Tuned Flan-t5-xl }  \\
 
      & {\textbf{Hit@1}} & {\textbf{Hit@5}} & {\textbf{Hit@1}}  & {\textbf{Hit@5}} & {\textbf{Hit@1}} & {\textbf{Hit@5}}    \\
      \midrule
      Baseline Method & 0.4575	&0.7406 & 0.4575	&0.7406 & 0.4575 &	0.7406\\
      MDG-MICL (0-shot) & \textbf{0.5} & \textbf{0.7877} & \textbf{0.4858} & \textbf{0.7689 } & 0.4434 & \textbf{0.7594}\\
      MDG-MICL (1-shot) & \textbf{0.4764} & \textbf{0.7972} & \textbf{0.4953} & \textbf{0.8113} & 0.467 & \textbf{0.75} \\
      MDG-MICL (2-shots) & \textbf{0.5047} & \textbf{0.8133} & \textbf{0.4858} & \textbf{0.816} & \textbf{0.5} & \textbf{0.7877}\\
      MDG-Cl & 0.434 & \textbf{0.7453} & \textbf{0.5} & \textbf{0.7642} & 0.4575 & \textbf{0.7547} \\
      MDG-MCQA & \textbf{0.5708} & \textbf{0.7547} & \textbf{0.5755} & \textbf{0.7453} & \textbf{0.5753} & \textbf{0.7642} \\
      DI-Cl & 0.3585 & 0.4717 & \textbf{0.4811} & 0.6934 & 0.3632 & 0.4953 \\
      DI-MCQA & \textbf{0.5519} & 0.5519 & \textbf{0.5708} & 0.5708 & \textbf{0.5519} & 0.5519 \\
      % \midrule
    \bottomrule
  \end{tabular}
  % \begin{tabular}{ p{3.8cm}p{1.3cm}p{1.3cm}p{1.3cm}p{1.3cm}p{1.3cm}p{1.3cm}p{1.3cm}}
  %   \toprule
  %   \multirow{2}{*}{Methods} &
  %     \multicolumn{2}{c}{Flan-t5-xl } &
  %     \multicolumn{2}{c}{Flan-t5-xxl } &
  %     \multicolumn{2}{c}{Tuned Flan-t5-xl }  \\
 
  %     & {\textbf{Hit@1}} & {\textbf{Hit@5}} & {\textbf{Hit@1}}  & {\textbf{Hit@5}} & {\textbf{Hit@1}} & {\textbf{Hit@5}}    \\
  %     \midrule
  %     Baseline Method & 0.4575	&0.7406 & 0.4575	&0.7406 & 0.4575 &	0.7406\\
  %     MG-DG (0-shot) & \textbf{0.5} & \textbf{0.7877} & \textbf{0.4858} & \textbf{0.7689 } & 0.4434 & \textbf{0.7594}\\
  %     MG-DG (1-shot) & \textbf{0.4764} & \textbf{0.7972} & \textbf{0.4953} & \textbf{0.8113} & 0.467 & \textbf{0.75} \\
  %     MG-DG (2-shots) & \textbf{0.5047} & \textbf{0.8133} & \textbf{0.4858} & \textbf{0.816} & \textbf{0.5} & \textbf{0.7877}\\
  %     MG-Cl & 0.434 & \textbf{0.7453} & \textbf{0.5} & \textbf{0.7642} & 0.4575 & \textbf{0.7547} \\
  %     MG-MCQA & \textbf{0.5708} & \textbf{0.7547} & \textbf{0.5755} & \textbf{0.7453} & \textbf{0.5753} & \textbf{0.7642} \\
  %     % \midrule
  %   \bottomrule
  % \end{tabular}
   \caption{\label{table:pretuned} This table shows the \textbf{Hit@5} and \textbf{Hit@1} scores achieved by the Baseline, MDG-MICL (0,1,2)-shots, MDG-Cl, MDG-MCQA, DI-Cl and DI-MCQA methods on Flan-T5-XL, Flan-T5-XXL and Tuned Flan-T5-XL models. The highlighted numbers indicate that the corresponding methods have outperformed the baseline method in terms of the \textbf{Hit@5}/\textbf{Hit@1} value. MDG-MICL consistently achieves the highest \textbf{Hit@5} and \textbf{Hit@1}, whereas DI-Cl and DI-MCQA have the worst \textbf{Hit@5}.}
\end{table}

\section{Discussion and Future work}
This paper proposes two different classes of methods, i.e., MDGM and DMGM, that leverage LLMs for solving the metadata to glossary matching problem.
MDGM methods use LLMs to generate good metadata descriptions, which we couple with similarity-based metrics for more refined matches. This class of methods is necessary when the glossary is likely to change during test time frequently, and repeated inferences are expensive. 
The second class of methods (DMGM) utilizes LLMs to infer which glossary items are potential matches directly. These methods are helpful when the metadata is too complex, and there is a significant difference between the description of the metadata and that of its closest glossary match.
Although we have shown that many of these methods can potentially obtain more accurate glossary matching, we can further improve them in several ways.
Our experiments show that DMGM methods perform poorly compared to MDGM methods. This may be due to the undesirable biases towards specific class labels that LLMs learned during training. One approach to mitigating these biases is using various calibration techniques \citep{xu2023knn,zhao2023llm,han2023prototypical}. Providing several positive and negative demonstrations in the classification and multiple-choice question-answer prompts may also help mitigate LLMs' default biases \citep{min2022rethinking}. 
We can further improve the MDGM methods that generate descriptions of metadata by constraining LLMs to sample words mainly from the glossary or providing LLMs with the top-$k$ glossary items and prompting LLMs to generate descriptions similar to those of the glossary items. This can be achieved by using the Constrained Beam Search algorithm \citep{hokamp-liu-2017-lexically} or simply manipulating the output distribution of LLMs before sampling such that it assigns higher weights to words from the glossary. It may also be helpful to use various prompt-tuning and prompt-editing methods \citep{zhang2023tempera} to further improve the efficiency of the prompts used with LLMs.
Although these directions remain intriguing, they warrant more in-depth empirical study, which we leave for future work.

\bibliography{references}

\end{document}